\documentclass[ reprint,
 amsmath,amssymb,
 aps,
]{revtex4-1}
\usepackage{graphicx}
\usepackage{subfigure}
\usepackage{float}
\usepackage{calrsfs} %$\mathcal{E}$
\usepackage{graphicx}% Include figure files
\usepackage{dcolumn}% Align table columns on decimal point
\usepackage{bm}% bold math
\begin{document}

\preprint{APS/123-QED}

\title{Electron current drive by fusion-product-excited lower hybrid drift instability}% Force line breaks with \\

\author{J. W. S. Cook$^1$}
\author{S. C. Chapman$^1$}%
\affiliation{$^1$Centre for Fusion Space and Astrophysics, Department of Physics, Warwick University, Coventry CV4 7AL, U.K.}
\author{R. O. Dendy$^{2,1}$}%

\affiliation{$^2$ Euratom/CCFE Fusion Association, Culham Science Centre, Abingdon, Oxfordshire OX14 3DB, U.K.}

\begin{abstract}
We present first principles simulations of the direct collisionless coupling of the free energy of fusion-born ions into electron current in a magnetically confined fusion plasma.  These simulations demonstrate, for the first time, a key building block of some ``alpha channelling''  scenarios for tokamak experiments.  A fully self-consistent electromagnetic 1D3V particle-in-cell code is used to evolve a parallel drifting ring-beam distribution of 3MeV protons in a 10keV thermal deuterium-electron plasma with realistic mass ratio. Collective instability gives rise to electromagnetic field activity in the lower hybrid range of frequencies. These spontaneously excited obliquely propagating waves undergo Landau damping on resonant electrons, drawing out an asymmetric tail in the distribution of electron parallel velocities, which carries a current. 

\end{abstract}

\maketitle

Optimal exploitation of the free energy of fusion products, for example the alpha-particles born at 3.5MeV in reactions between thermal deuterons and tritons at ~10-20keV, is central to achieving fusion power through magnetic confinement of plasma. In the traditional framework, this energy is transferred through multiple collisions to the thermal electrons on a cumulative timescale $\sim$1s; electron heating by fusion alpha-particles has been observed in the TFTR\cite{1} and JET\cite{2} tokamaks. The electrons in turn sustain the temperature of the thermal ions to which they are collisionally coupled. It may be preferable, however, to use some of this fusion product  free energy in “alpha channelling” scenarios. This term, coined by Fisch and Rax\cite{3}, refers to mechanisms by which rapid collisionless collective instabilities (natural or induced) of the fusion product population could directly benefit the plasma equilibrium, for example by helping sustain toroidal current\cite{4,5,6}. Here we report particle-in-cell (PIC) simulations of fusion-born protons in deuterium plasmas that demonstrate from first principles, for the first time, a key alpha channelling phenomenon for tokamak fusion plasmas. We focus on the collective instability of centrally born fusion products on banana orbits at the plasma edge, a population known to be responsible for observations of ion cyclotron emission in JET\cite{7} and TFTR\cite{8}. A fully self-consistent electromagnetic 1D3V PIC code evolves a parallel drifting ring-beam distribution of 3MeV protons in a 10keV thermal deuterium-electron plasma with realistic mass ratio. Collective instability gives rise to electromagnetic field activity in the lower hybrid range (LH) of frequencies.
The spontaneously excited obliquely propagating waves undergo Landau damping on resonant electrons, drawing out an asymmetric tail in the distribution of electron parallel velocities, which carries a current. These simulations thus demonstrate a key building block of some alpha channelling scenarios: the direct collisionless coupling of fusion product free energy into a form which can help sustain the basic equilibrium of the tokamak plasma.

Spatially localized inversions of the velocity distribution of fusion-born ions can arise due to the particle energy and pitch angle dependence of particle orbits. Alpha-particles born with pitch angles just inside the trapped-passing boundary generate ion cyclotron emission\cite{7,8}, for example. This motivates our model\cite{9} of the initial fusion product velocity distribution function as a ring travelling anti-parallel to the magnetic field with distribution function $f_{p} = 1/(2\pi v_r) \delta(v_{\vert\vert} - u) \delta(v_{\perp} - v_{r})$; $u$ is the magnetic field aligned velocity and $v_r$ is the perpendicular velocity. The radial extent of the emitting region is of order a few energetic particle gyroradii in JET and TFTR\cite{7,8}.

Our simulations use physical mass ratios to ensure that the physically relevant instability is excited. We select parameter values similar to those in the edge of a large tokamak, subject to computational resource constraints. The minority energetic fusion product protons at 3MeV have pitch angle $135^\circ$ and their number density $n_p$ is $1\%$ of that of the background (Maxwellian) 10keV deuterons $n_d$. The electron number density of the initially quasineutral plasma is $n_e = 10^{18}m^{-3}$ and the electron beta $\beta_e=\beta_{d} = 3 \times 10^{-4}$. The speed of the energetic protons is approximately half the Alfv\'{e}n speed $V_A$ given the applied magnetic field $B = 3T$. These parameter values imply a total energy of the energetic protons $\sim\!1.7$ times that of the thermal deuterons and electrons combined. While this is $\sim\!10$ times the value anticipated for next step fusion plasmas\cite{10}, it is necessary so as to drive the instability on an acceptable timescale computationally.

The collective instability of the energetic protons can be identified as a form of lower hybrid drift instability (LHDI), a topic of widespread relevance to space and astrophysical plasmas\cite{11,12}, while LH waves are used for current drive\cite{13,14,15} in the fusion context and are associated with acceleration mechanisms in ionospheric plasmas\cite{16}. The treatment here extends LHDI theory in several respects: the energetic ion population does not contribute to the plasma equilibrium, unlike most space and astrophysical applications, where the ion beams are typically associated with currents and gradients central to the equilibrium; values of the key dimensionless parameters are guided by large tokamak edge plasma conditions; and the physically correct mass ratio of electrons to ions is used, which was not possible in some of the classic computational studies of LHDI.

Our fully self-consistent, relativistic kinetic simulations use a particle-in-cell code
epoch1d based on the approach of Ref.\cite{17}. Computational macroparticles represent the particle distribution functions in full three dimensional velocity space. All three components of all vector quantities, that is, particle velocities and electromagnetic fields, are functions of a single spatial coordinate ($x$, referred to as the direction of variation or the simulation domain) and  time $t$. Field and particle boundary conditions are  periodic. Wavevectors are then parallel or antiparallel to the $x$ direction. To focus on obliquely propagating modes, we set the background field  at an angle $\theta = 84^\circ$ to the $x$ direction. The fields and bulk plasma parameters are resolved (in $x$) by computational cells of width $\Delta x = \lambda_D/10$, where $\lambda_D$ is the electron Debye length. The simulation domain (in $x$) is optimized to capture the rather broad range of characteristic lengthscales of the energetic protons, background deuterons and electrons. Each of the  $N_G = 2048$ cells corresponds to species  gyroradii $\rho$ of  $\sim\!2-3 \rho_{p}$, $\sim \!10 \rho_{d}$ and $\sim\! 1000 \rho_{e}$. The simulation requires over 2 million computational particles to give good phase space resolution and is typically iterated over $\sim\!8\times 10^5$ timesteps.

\begin{figure}[h]
 \begin{center}
    \includegraphics[width=0.48\textwidth]{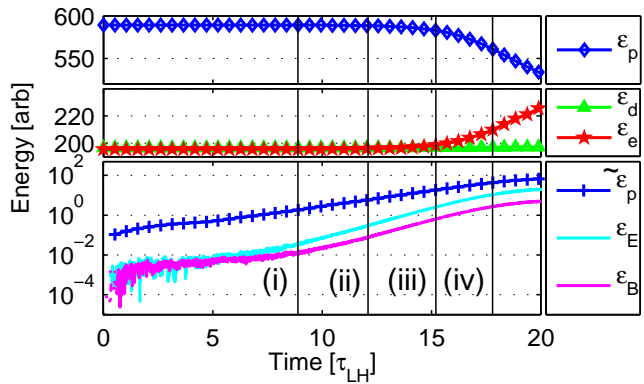}
  \end{center}
\begin{small}
 \vspace{-0.03\textheight}
\caption{Time evolution of: the total kinetic energy $\varepsilon_i$ in each plasma species; the energies in the electric (magnetic) field $\varepsilon_E$ ($\varepsilon_B$); the proton fluctuation energy ${\tilde{\varepsilon}}_p$. }
\end{small}
\end{figure}

An overview of the energy flows in the simulation is shown in Figure 1. The total electric and magnetic energy in excited fields in the simulation are obtained by summing the (suitably normalized) energy densities over all grid cells giving electric  $\varepsilon_E=\sum E_i^2$ and magnetic $\varepsilon_B=\sum (B_i^2-B_0^2)$   field energies  where $B_{0}$ is the applied magnetic field. Both the electric field energy $\varepsilon_E$ (light blue trace) and the magnetic field energy $\varepsilon_B$ (magenta trace) rise  with time. Four stages of the evolving simulation are indicated by vertical lines (i)-(iv) corresponding to times $8.9 , 12.0, 15.2$ and $17.8$ lower hybrid oscillation periods $\tau_{LH}$ respectively. Stage (i) corresponds to the onset of the linear phase of field growth, which is well developed by stages (ii) and (iii). By stage (iv) the wave amplitude is approaching its saturated level.  The magnetic fluctuations contain nearly an order of magnitude less energy than the electric fluctuations, implying that the waves produced by the instability are largely electrostatic.
 
The total kinetic energy of each particle species is obtained by summing over all the computational particles in the system. From stage (iii) onwards, Figure 1 shows that the total kinetic energies of the protons and electrons approximately mirror each other, the protons losing energy as the electrons gain energy, whilst the deuterons show little change. 
 
 \begin{figure}[ht]
\includegraphics[width=0.5\textwidth]{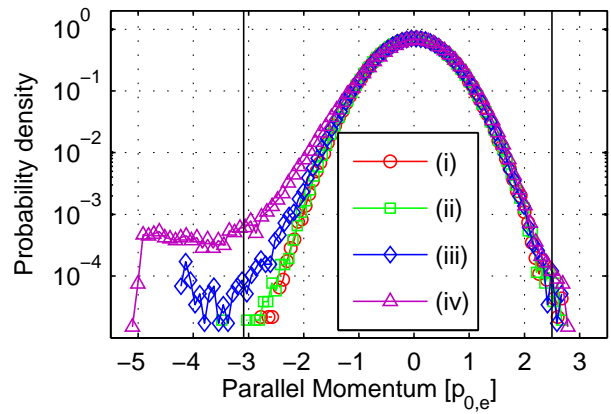}%[trim = 1cm 6cm 0cm 9cm, clip, width=7.5cm]
\begin{small}
 \vspace{-0.03\textheight}
\caption{Electron parallel momentum distribution function at four snapshots in time (i)$-$(iv) (see Figure 1) in units of initial electron $rms$ momentum $p_{0,e}$. Vertical solid black traces indicate the phase velocities of the fundamental modes shown in Figure 3, in non-relativistic electron momentum space.}
\end{small}
\end{figure}
 
To  obtain the relative gain, or loss, in the energy of each population we  define a change in total kinetic energy of the ensemble of particles $\Delta \varepsilon_{i} = <\!\varepsilon_i(t)\!> - <\!\varepsilon_i(t=0)\!\!> $,  and a total kinetic energy of fluctuations $\tilde{\varepsilon}_{i}(t)=<\!|\varepsilon_j(t) - <\!\varepsilon_i(t=0)\!>\!|\!>$ relative to the initial conditions, where $<\!...\!>$ indicates an average over all the particles of species $i$. 
The time variation of the proton fluctuation energy $\tilde{\varepsilon}_p$ is shown in Figure 1. It grows from the start of the simulation, ultimately increasing by three orders of magnitude, whereas the total proton kinetic energy declines by much less than one order of magnitude. This reflects the role of proton energy dispersion in the early phase of the instability. The electron kinetic energy $\varepsilon_e$ rises with the electric field energy $\varepsilon_E$ during the linear stage of the simulation, arising from electron participation in the principally electrostatic waves excited by the instability. The corresponding effect for the deuterons is also visible, but is much less due to their higher mass.

\begin{figure}[h]
 \begin{center}
    \includegraphics[width=0.48\textwidth]{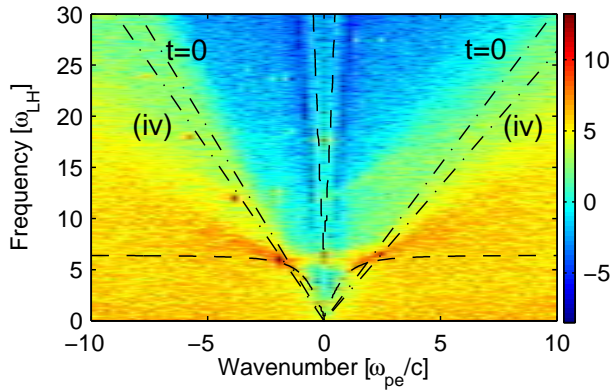}%[trim = 1cm 6cm 0cm 9cm, clip, width=7cm]
  \end{center}
\begin{small}
 \vspace{-0.03\textheight}
\caption{Fast Fourier transform of the $E_x$ electric field over the spatial domain and the time interval $10 \leq t/\tau_{LH} \leq 18$. Colour denotes field power in arbitrary units. The electron-deuteron cold dispersion relation (black dash trace) and the positions of the $v_x$ proton distribution function peaks at $t\!=\!0$ and time (iv) (black dash-dot traces) are overplotted.}
\end{small}
\end{figure}

We now turn to  the evolving distribution function of the electron parallel momentum $p_{\parallel}$  shown in Figure 2 for times (i)-(iv). From stage (iii) onwards the electron distribution function develops an asymmetric tail in $p_{\parallel}$ reflecting net directional electron acceleration. We infer Landau damping of the excited waves on resonant electrons which results in the flattening of the negative tail of the $p_{\parallel}$ distribution function.

This is confirmed by the analysis shown in Figure 3 where we plot the spatio-temporal fast Fourier transform of the electric field component along the simulation direction  $E_x(\omega,k)$, transformed from $E_x(x,t)$ for the time interval $10 \leq t/\tau_{LH} \leq 18$ ending at time (iv). The oblique cold-plasma normal mode of the background deuteron-electron plasma is marked by the black dashed trace. The presence of the additional energetic proton population  modifies the nature and dispersion properties of the normal modes of the  plasma and, through resonance, couple energy to these modified modes. The Figure shows peaks in intensity at $(\omega,k ) \simeq (6\omega_{LH},\pm 2 \omega_{pe}/c)$  which indicate  resonance of the  proton population with a modified deuteron-electron normal mode branch that lies on the surface between the lower extraordinary wave and the whistler wave in $\omega,k_{\perp},k_{\parallel}$ space. The locations in $\omega,k$ space where coupling takes place correspond to where the velocities of the peaks of the proton distribution function in $v_x$ approximately match the phase velocities of the normal mode.  As the proton distribution function evolves, the positions in $v_x$ of the peaks in the distribution function move. These velocities at $t=0$ and at time (iv) are plotted on Figure 3 as black dash-dot traces. We note that there are harmonics of the dominant backward propagating (in negative $x$ direction) mode at $(\omega,k) \simeq (6\omega_{LH},-2 \omega_{pe}/c)$.

\begin{figure}[]
\includegraphics[width=0.48\textwidth]{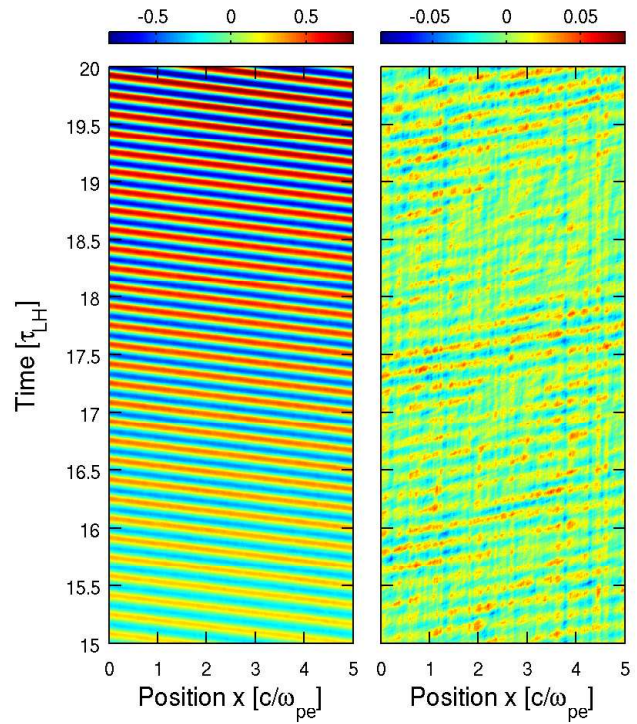}
\begin{small}
 \vspace{-0.03\textheight}
\caption{Spatio-temporal plots of the backward $E_x^-(x,t)$ (left) and forward $E_x^+(x,t)$ (right) travelling wave components of $E_x(x,t)$. Colour indicates normalized field amplitude.}
\end{small}
\end{figure}

On Figure 2 we indicate with vertical lines the two electron parallel momenta resonant with the dominant backward and forward electromagnetic structures shown in Figure 3.
Taken together, Figures 2 and 3 indicate that the electrons are principally accelerated by the dominant wave at negative phase velocity,  which is in turn
excited by the backward-drifting fusion products.

To capture the coherent oscillatory features of the dominant excited fields, and hence their phase resonant characteristics, we decompose the electric field $E_x(x,t)$ into component waves travelling with positive and negative phase velocities. These are $E_x^+(x,t)=IFT[E_x(\omega\!>\!0,k\!>\!0)+E_x(\omega\!<\!0,k\!<\!0)]$ and $E_x^-(x,t)=IFT[E_x(\omega\!<\!0,k\!>\!0)+E_x(\omega\!>\!0,k\!<\!0)]$, where IFT denotes the inverse Fourier transform, which we plot in Figure 4. The sum of $E_x^+(x,t)$ and $E_x^-(x,t)$  recreates the original field.  In Figure 4 , the amplitude of the  backward travelling wave is approximately an order of magnitude greater than the  forward travelling wave and  is dominated by a single long wavelength component. The parallel phase velocity of this backward travelling wave is shown by the vertical line on the left of Figure 2.

\begin{figure}[]
\includegraphics[width=0.48\textwidth]{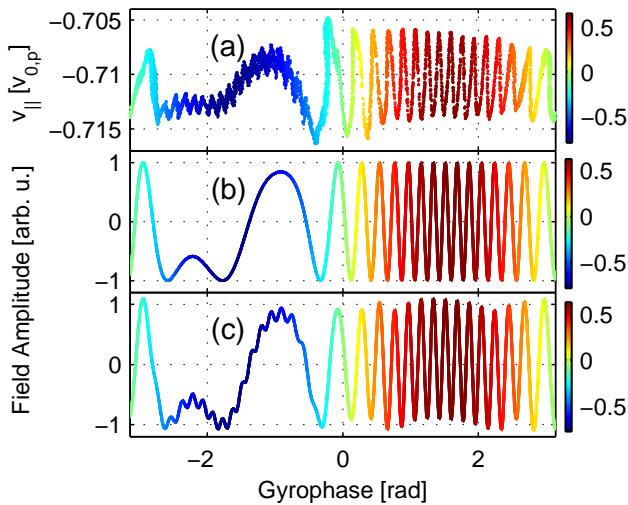}
\begin{small}
 \vspace{-0.03\textheight}
\caption{Panel (a): Snapshot at time (iii) of the energetic proton velocity space.  Velocity space coordinates are $v_{\parallel}$ (abscissa), gyrophase (ordinate) and $v_x$ (colour).   Panel (b): Normalized wave amplitude seen by protons at resonance with the dominant wave plotted as a function of phase space (see text). Panel (c): As in panel (b) for a sum of dominant wavemode and the counter-propagating damped mode.}
\end{small}
\end{figure}

The predominantly drift character of the instability is directly seen in  the fully resolved velocity space of the PIC simulation. The velocity of each particle can be expressed in terms of a field aligned component $v_{\parallel}$, a component aligned with the simulation domain (and $k$) $v_x$, and gyrophase $arctan(v_{\perp,1}/v_{\perp,2})$. We test for velocity space patterns in these coordinates at snapshot (iii) when the wavefields are well established. We select protons from a narrow region in configuration space $\delta x$ that is smaller in extent that the wavelength of the dominant wavemode ($\delta x/\lambda = 0.1$). The  coordinates $v_{\parallel}$,  $v_x$, and the gyrophase  for each particle are plotted in panel (a) of Figure 5, where gyrophase is along the abscissa, $v_{\parallel}$ is along the ordinate and colour indicates $v_x$. We see asymmetry in the oscillatory pattern in $v_{\parallel}$ which is a function of $v_x$: a characteristic of drift instability rather than gyroresonance. As such, these oscillatory perturbations in $v_{\parallel}$ should vary as the wave amplitude at the point of resonance $v_x=\omega/k$,  and in panel (b) of Figure 5 we confirm that there is indeed resonance with the dominant LH wave identified above. Panel 5(b) plots on the ordinate the normalized wave amplitude $\mathcal{E}$, that is, the sine of the wave phase $(kx_R-\omega t_R)$ at the location $x_R$ and time $t_R$ of unperturbed test particles that satisfy the condition for resonance with the dominant wave $(\omega, k)$. The point of resonance $(x_R,t_R)$ is obtained for test particles which initially are distributed uniformly in gyrophase $\phi$ and which  follow unperturbed cycloidal orbits. To reconstruct a single snapshot in $x$ and $t$ (to compare with the simulation snapshot of panel (a)) these test particle orbits are then advanced in $x$, $v_x$ and $\phi$ to a single point $(x_1,t_1)$. We plot in panel (b), for a snapshot in space and time $(x_1,t_1)$, the normalized wave amplitude experienced by the particle at the point of resonance $\mathcal{E}(x_R,t_R)$ against particle gyrophase $\phi(t_1)$ with $v_x(\phi(t_1))$, represented in colour. We  see that panel 5(b) closely tracks the large scale oscillations seen in the PIC code velocity space of panel 5(a). We refine this idea in panel 5(c) where we plot the effective wave amplitude arising from both the forward and backward waves identified in Figure 4; the weaker second wave can be seen to introduce fine scale oscillations visible in panel 5(a).

These first principles simulations demonstrate, for the first time, key physical elements of alpha channelling scenarios for future tokamak plasmas: the excitation of LH waves by a fusion product population, whose functional form and parameters are aligned with prior observations on JET and TFTR, combined with the subsequent Landau damping of the excited waves on resonant electrons, drawing out an asymmetric tail in their parallel velocity distribution, which carries a current. These simulations also contribute to the question of what instabilities may arise –- transiently, locally, or otherwise –- during the initiation and propagation of fusion burn in tokamak plasmas. Furthermore they deepen our understanding of LH drift instabilities, which may be widespread in space and astrophysical plasmas, by extending their study into parameter ranges that approximate (subject to computational resource constraints) edge plasma conditions in large tokamaks.
 
\acknowledgements  The authors acknowledge the EPSRC for support. This work was supported in part by EURATOM under a contract of association with the CCFE. The views expressed do not necessarily reflect those of the European Commission.

\end{document}